\documentclass[prb,twocolumn,showpacs,preprintnumbers,amsmath,amssymb]{revtex4}
\usepackage{graphicx}
\usepackage{dcolumn}
\usepackage{bm}

\preprint{submitted to PRB}

\begin{document}

\title{Appearance of Half-Metallicity in the Quaternary Heusler Alloys}

\author{Iosif Galanakis}\email{I.Galanakis@fz-juelich.de}

\affiliation{Institut f\"ur Festk\"orperforschung, Forschungszentrum
  J\"ulich, D-52425 J\"ulich, Germany}

\date{\today}

\begin{abstract}
 I report systematic first-principle calculations of the quaternary
 Heusler alloys like Co$_2$[Cr$_{1-x}$Mn$_x$]Al, Co$_2$Mn[Al$_{1-x}$Sn$_x$] and
 [Fe$_{1-x}$Co$_x$]$_2$MnAl. I show
 that when the two limiting cases (x=0 or 1) correspond to a half-metallic
 compound,
 so do the intermediate cases. Moreover the total spin moment $M_t$ in $\mu_B$ scales
 linearly with the total number of valence electrons $Z_t$ (and thus with the
 concentration $x$) following the relation $M_t=Z_t-24$,
 independently of the origin of the extra valence electrons, confirming the Slater-Pauling
  behavior of the normal Heusler alloys. Finally I discuss in all cases the trends in
   the atomic projected DOSs and in the atomic spin moments.

\end{abstract}


\maketitle

\section{Introduction}

Heusler alloys consist a large family of intermetallic compounds
which attract regularly considerable attention due to the variety
of magnetic phenomena which they present. Lately the interest has
been focused  on the  ones being half-metals (HM)\footnote{I will
use the abbreviation HM to denote both the ``half-metal'' and the
``half-metallic''.} like NiMnSb or
Co$_2$MnSn.\cite{deGroot83,Galanakis00,Galanakis02a,Galanakis02b}
These compounds are ferromagnets
   for which the minority spin-band presents a gap and the Fermi level falls within
   this gap. Thus the spin polarization at the Fermi level is 100\% and these materials
    are of special interest for spintronic
    applications;\cite{Soulen98,Prinz98} the spin of the electron and
    not the  charge is used as the property to control the device.
Contrary to other HM systems like the diluted magnetic
semiconductors or the manganites and some oxides,\cite{Soulen98}
the half- and full-Heusler alloys (e.g. NiMnSb and Co$_2$MnSn,
   respectively) and the zinc-blende compounds like CrAs\cite{GalanakisCrAs}
   present very high Curie temperatures making them
   attractive for industrial applications.

Lately I have devoted a lot of work on the latter compounds trying
to address three main points i) the origin of the
half-metallicity\cite{Galanakis02a,Galanakis02b,GalanakisCrAs} ii)
whether the half-metallicity is preserved at the surfaces or
films\cite{GalanakisCrAs,iosifSURF} and iii) to predict new
half-metallic materials which may present better growth on top of
semiconductors.\cite{Galanakis02a,Galanakis02b,GalanakisCrAs}
Using first principle calculations it was shown in the case of the
ordered half-Heusler alloys like NiMnSb that the gap is formed
between the occupied bonding $d$ states resulting from the
interaction between the higher- and the lower-valent transition
metal atoms and the corresponding antibonding
states.\cite{Galanakis02a} For the full-Heusler alloys like
Co$_2$MnSn the situation is more complicated since there are also
states located only at the Co sites and the resulting gap is
tiny.\cite{Galanakis02b} The total spin moment for the latter
compounds is given by the relation $M_t$=$Z_t$-24, where $M_t$ is
the total spin moment in $\mu_B$ and $Z_t$ the total number of
valence electrons. I will discuss this behavior in detail in a
later paragraph.

In the present contribution I will conclude this study by
examining the behavior
 of the so-called quaternary Heusler alloys. In the latter compounds, one of the four
  sites is occupied by two different kinds of  neighboring elements like Co$_2$[Cr$_{1-x}$Mn$_{x}$]Al
  where the Y site is occupied by Cr or Mn atoms (see Fig. 1 in Ref. \onlinecite{Galanakis02b}
   for the definition of the structure). To perform this study I used the
Korringa-Kohn-Rostoker (KKR) method\cite{Papanikolaou02} developed
  by Professor Akai, which has been already used with success to study the magnetic
   semiconductors.\cite{Akai98} The atomic disorder at a specific site is implemented using the
coherent potential approximation (CPA).
   The space is divided in non-overlapping muffin-tin spheres and the vanishing charge in the
   interstitial region is considered to be constant and charge neutrality is imposed.
   Such an approximation is reasonable to describe a close-packed structure like the one of
   the Heusler alloys. To check the validity of this description I compared the
   total spin moment with the full-potential calculations performed in Ref.
   \onlinecite{Galanakis02b}. For all the compounds with the exception of Co$_2$VAl
   both methods produced similar magnetic properties for the
   experimental lattice constants.\cite{lattice-csts} V $3d$
   wavefunctions have a large extent and the muffin-tin
   approximation is not adequate to describe the V magnetism; thus I had
to use a lattice constant 4\% larger than the experimental one to
get a half-metallic compound, as predicted by the full-potential
KKR for the experimental lattice constant. I assumed that the
lattice constant varies linearly with the concentration $x$ which
has been verified for several quaternary
alloys.\cite{lattice-csts} Finally I should mention that the total
spin moment is not exactly an integer due to numerical
inaccuracies. To decide whether an alloy is half-metallic or not I
used the total DOS as a criterion, as was also the case in the
previous studies.\cite{Galanakis02a,Galanakis02b,GalanakisCrAs} I
will firstly focus my study on the behavior of the total spin
moment for several cases and show that it follows the
Slater-Pauling behavior when the limiting cases are half-metals.
Afterwards I will study the case where the Y site (see Fig.1 in
Ref. \onlinecite{Galanakis02b}) is occupied by two different
elements like Co$_2$[Cr$_{1-x}$Mn$_x$]Al, followed by the case
when there are two different types of $sp$ atoms like
Co$_2$Mn[Al$_{1-x}$Sn$_x$]. Final case is when the X sites are
disordered like [Fe$_{1-x}$Co$_x$]$_2$MnAl. At the end I summarize
and conclude.

\begin{figure}[t]
\begin{center}
\includegraphics[scale=0.52]{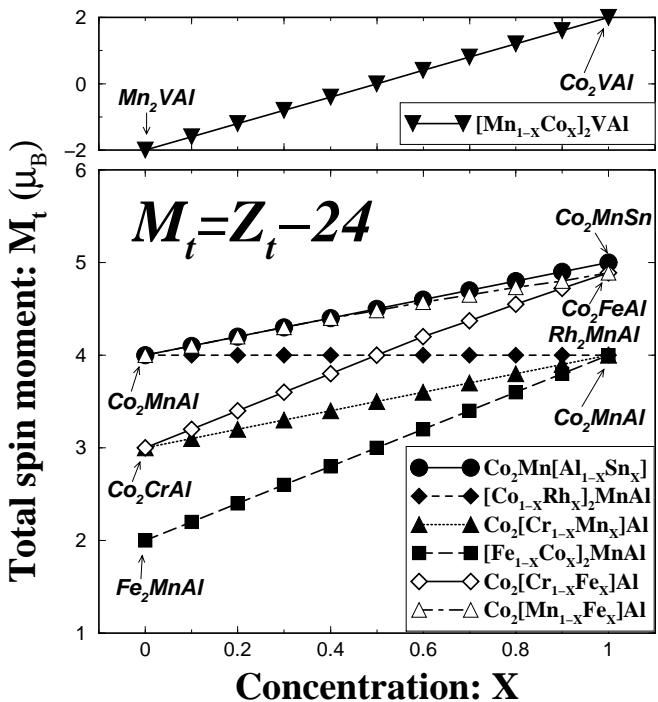}
\end{center}
\caption{Calculated total spin moment $M_t$ in $\mu_B$ for a
variety of compounds as a function of the concentration $x$
($x$=0,0.1,0.2,...,0.9,1). I assumed that the lattice constant
varies linearly with the concentration $x$. With filled
geometrical objects the cases obeying the rule $M_t=Z_t-24$ where
$Z_t$ is the total number of valence electrons; this is the
so-called Slater-Pauling behavior.} \label{fig1}
\end{figure}

\begin{table}
\caption{\label{Table1}Atom-projected spin moments for
Co$_2$[Cr$_{1-x}$Mn$_x$]Al with $x$=0,0.1,0.2,...,0.9,1. The spin
moments have been scaled to one atom. The total moment is given by
the relation $m^\mathrm{total}=2\times
m^\mathrm{Co}+(1-x)m^\mathrm{Cr}+ x
m^\mathrm{Mn}+m^\mathrm{Al}+m^\mathrm{int}$, where $\mathrm{int}$
stands for the interstitial region.}
\begin{ruledtabular}
\begin{tabular}{lrrrrr}
Co$_2$[Cr$_{1-x}$Mn$_x$]Al & $m^\mathrm{Co}$ & $m^\mathrm{Cr}$
&$m^\mathrm{Mn}$ & $m^\mathrm{Al}$ & $m^\mathrm{Total}$ \\ \hline
Co$_2$CrAl & 0.698 &  1.686 & -- & -0.060 & 3.007  \\
Co$_2$[Cr$_{0.9}$Mn$_{0.1}$]Al & 0.709 &  1.615 & 3.116 & -0.060
&3.106\\ Co$_2$[Cr$_{0.8}$Mn$_{0.2}$]Al & 0.718 & 1.547 & 3.068 &
-0.061 & 3.206 \\ Co$_2$[Cr$_{0.7}$Mn$_{0.3}$]Al & 0.722 & 1.483
&3.020 &-0.061 &3.305\\ Co$_2$[Cr$_{0.6}$Mn$_{0.4}$]Al &  0.726&
1.416 & 2.971  & -0.061 & 3.405 \\ Co$_2$[Cr$_{0.5}$Mn$_{0.5}$]Al
& 0.731 & 1.342 & 2.920 & -0.062 & 3.505\\
Co$_2$[Cr$_{0.4}$Mn$_{0.6}$]Al & 0.729& 1.317 & 2.864& -0.063
&3.605\\ Co$_2$[Cr$_{0.3}$Mn$_{0.7}$]Al & 0.738& 1.220 &2.806 &
-0.063 & 3.706 \\ Co$_2$[Cr$_{0.2}$Mn$_{0.8}$]Al & 0.746&1.093  &
2.744 & -0.064 & 3.806 \\ Co$_2$[Cr$_{0.1}$Mn$_{0.9}$]Al & 0.752 &
0.834 & 2.682 & -0.064 & 3.907 \\ Co$_2$MnAl & 0.755 & -- & 2.603
& -0.064 &  4.020
\end{tabular}
\end{ruledtabular}
\end{table}

As it was shown in Ref. \onlinecite{Galanakis02b} the total spin
moment in the case of the half-metallic full-Heusler alloys obeys
the rule $M_t=Z_t-24$, where $M_t$ is the total spin moment in
$\mu_B$ and $Z_t$ the total number of valence electrons. The 24
arises from the fact that there are in total 12 occupied minority
states; \textit{e.g.} in the case of Co$_2$MnAl Al provides one
minority $s$ band very low in energy and three minority $p$ bands
which accommodate also transition metal $d$ electrons. There are
also five occupied bonding $d$ bands created from the interaction
between the Mn and the Co atoms. The Fermi level falls within the
gap created by the occupied triple degenerated $d$ states and the
unoccupied double degenerated $d$ states which are exclusively
located at the Co sites and are permitted by the symmetry of the
crystal. The moments behavior is the well-known Slater-Pauling
behavior known from the  binary alloys. In the latter compounds
the spin moment decreases with $Z_t$ since the spin-up states are
completely occupied and the extra electrons occupy spin-down
states reducing the total spin moment. In the case of the full
Heusler alloys the Fermi level is fixed within the minority band
gap and the extra electrons as we change the chemical elements
occupy exclusively spin-up states and the total spin moment
increases. Using the KKR-CPA method we calculated the total spin
moment for several quaternary alloys taking into account several
possible combinations of chemical elements and assuming in all
case a concentration increment of 0.1. I resume my results in Fig.
\ref{fig1}. The first possible case is when I have two different
low-valent transition metal atoms at the Y site like
Co$_2$[Cr$_{1-x}$Mn$_x$]Al. The total spin moment varies linearly
between the 3 $\mu_B$ of Co$_2$CrAl and the 4 $\mu_B$ of
Co$_2$MnAl. In the case of the Co$_2$[Cr$_{1-x}$Fe$_x$]Al and
Co$_2$[Mn$_{1-x}$Fe$_x$]Al compounds and up to around $x$=0.6 the
total spin moment shows the SP behavior but for larger
concentrations it slightly deviates to account for the not integer
value of Co$_2$FeAl.\cite{Galanakis02b} This behavior on the
figure is clearly seen when we compare the lines for the
Co$_2$[Mn$_{1-x}$Fe$_x$]Al and Co$_2$Mn[Al$_{1-x}$Sn$_x$]
compounds; the latter family following the SP behavior. The second
case is when one mixes the $sp$ elements, but as I just mentioned
these compounds also obey the rule for the total spin moments. The
third and final case is to mix the higher valent transition metal
atoms like in [Fe$_{1-x}$Co$_x$]$_2$MnAl and
[Rh$_{1-x}$Co$_x$]$_2$MnAl alloys. In the first case the total
spin moment varies linearly between the 2 and 4 $\mu_B$ of
Fe$_2$MnAl and Co$_2$MnAl compounds, respectively. Rh is
isoelectronic to Co and for the second family of compounds we find
a constant integer value of 4 $\mu_B$ for all the concentrations.
A special case is Mn$_2$VAl which has less than 24 electrons and
the total spin moment is -2 $\mu_B$. If now I mix Mn and Co I get
a family of compounds where the total spin moment varies linearly
between the -2 $\mu_B$ and the 2 $\mu_B$ and  for $x$=0.5 I get
the case of a paramagnetic compound consisting of magnetic
elements. Thus all the compounds obey the rule $M_t$=$Z_t$-24,
showing the Slater-Pauling behavior regardless of the origin of
the extra charge. In the next paragraphs I will analyze in detail
every case.

\begin{figure}[t]
\includegraphics[scale=0.5]{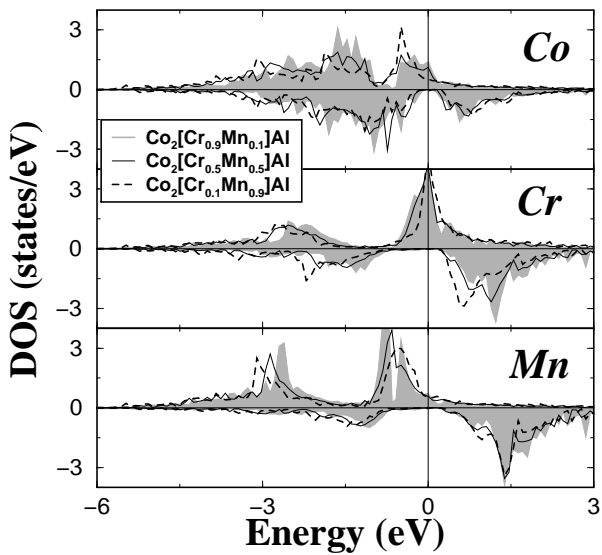}
 \caption{\label{fig2}Atom-projected DOS of Co$_2$[Cr$_{1-x}$Mn$_x$]Al for
 three different values of $x$: 0.1,0.5,0.9. The DOS's have been scaled to one atom. }
\end{figure}

The first large family of quaternary alloys which are susceptible
to be HMs are the ones where there are two kinds, Y and Y$^\star$,
of low-valent transition metal atoms :
X$_2$(Y$_{1-x}$Y$^\star_x$)Z, where X stands for the high-valent
transition metal atom and Z for the $sp$ atom. I will concentrate
my analysis on Co$_2$[Cr$_{1-x}$Mn$_x$]Al. Already in Ref.
\onlinecite{Galanakis02b} it was shown that both Co$_2$CrAl and
Co$_2$MnAl are HMs with a total spin moment of 3 $\mu_B$ and 4
$\mu_B$, respectively. In Fig. \ref{fig2} I present the
atom-projected DOS and in Table \ref{Table1} the spin moments
scaled to one atom for several concentrations. For all
concentrations, one gets a HM system since the Fermi level falls
within the minority gap. The width of the gap is the same for all
compounds since the gap is formed between states located
exclusively at the Co sites and which are little affected by the
lower-valent transition metal atoms like Mn and
Cr.\cite{Galanakis02b} This is clearly seen in Fig. \ref{fig2}
where the gap in the case of the Cr and Mn atoms is much larger
than for the Co atoms.
 The total spin moment scales linearly
between the two extremes reflecting the half-metallicity. For low
Mn concentrations, the Mn atom acts like an impurity and the $e_g$
and $t_{2g}$ electrons are well separated. This is clearly seen in
the Mn DOS, where for small Mn concentration there is a double
peak in the majority band just below the Fermi level. The lower
peak is the $e_g$ electrons and the higher peak the $t_{2g}$
electrons of Mn. When the Mn concentration increases these states
overlap and they can not be any more distinguished and Mn shows a
more itinerant-like magnetism. The states lower in energy around
-3 eV are $t_{2g}$-like states which couple to the $p$ states of
Al and fill the bands created by the latter ones. In the case of
Cr the behavior is the inverse. It is for the case of high Cr
concentrations, where the $e_g$ and $t_{2g}$ are distinguished.
For low Cr concentrations the population of the occupied minority
states is larger and they are lower in energy than for high
concentrations attracting also the unoccupied states lower in
energy and resulting to a smaller spin moment. The differences in
the Co DOS mainly arise from the different position of the Cr and
Mn majority bands through the Coulomb interaction and the Co spin
moment only slightly changes with the concentration. Finally in
the case of Co$_2$[Cr$_{1-x}$Fe$_x$]Al and
Co$_2$[Mn$_{1-x}$Fe$_x$]Al compounds the situation is similar but
the Co moments show more important changes since Fe can not
account only by itself for the extra electron and the Co moment
has also to reach a higher value. This was discussed in detail in
Ref. \onlinecite{Galanakis02b}.

\begin{table}
\caption{\label{Table2} Same as Table I for the
Co$_2$Mn[Al$_{1-x}$Sn$_x$] compounds.}
\begin{ruledtabular}
\begin{tabular}{lrrrrr}
Co$_2$Mn[Al$_{1-x}$Sn$_x$] & $m^\mathrm{Co}$ & $m^\mathrm{Mn}$
&$m^\mathrm{Al}$ & $m^\mathrm{Sn}$ & $m^\mathrm{Total}$ \\ \hline
Co$_2$MnAl & 0.755 & 2.603 & -0.064 & -- &  4.020   \\
Co$_2$Mn[Al$_{0.9}$Sn$_{0.1}$] & 0.766 & 2.673 & -0.065 & -0.047 &
4.113 \\ Co$_2$Mn[Al$_{0.8}$Sn$_{0.2}$] & 0.789 & 2.744 & -0.067 &
-0.048 & 4.223 \\ Co$_2$Mn[Al$_{0.7}$Sn$_{0.3}$] & 0.800 & 2.821 &
-0.068 & -0.049 & 4.324 \\ Co$_2$Mn[Al$_{0.6}$Sn$_{0.4}$] & 0.820
& 2.889 & -0.069 & -0.050 & 4.434 \\
Co$_2$Mn[Al$_{0.5}$Sn$_{0.5}$] & 0.841 & 2.951 & -0.069 & -0.051 &
4.541 \\ Co$_2$Mn[Al$_{0.4}$Sn$_{0.6}$] & 0.863 & 3.010 & -0.070 &
-0.051 & 4.644  \\ Co$_2$Mn[Al$_{0.3}$Sn$_{0.7}$] & 0.882 & 3.069
& -0.070 & -0.052 & 4.743 \\ Co$_2$Mn[Al$_{0.2}$Sn$_{0.8}$] &
0.902 & 3.125 & -0.070 & -0.052 & 4.843 \\
Co$_2$Mn[Al$_{0.1}$Sn$_{0.9}$] & 0.925 & 3.180& -0.071 & -0.052 &
4.946 \\ Co$_2$MnSn & 0.944 & 3.235 & -- & -0.052 & 5.043
\end{tabular}
\end{ruledtabular}
\end{table}

Now I will go on with the case of the X$_2$Y[Z$_{1-x}$Z$^\star_x$]
compounds, where I change the charge at the Z site. I studied both
the Co$_2$Mn[Al$_{1-x}$Si$_{x}$] and Co$_2$Mn[Al$_{1-x}$Sn$_{x}$].
Si and Sn are isoelectronic and both families present the same
behavior, thus I will restrict my presentation to the second
family. The moment changes from 4 up to 5 $\mu_B$ linearly and
thus all the intermediate cases are HMs. In Table \ref{Table2} I
have gathered the atom-resolved spin-moments. The $sp$ atoms show
a practically constant moment and the extra charge is taken into
account by the transition metal atoms. If I look carefully at the
Mn majority spin band (not presented here), there are unoccupied
states at the vicinity of the Fermi level which pass under the
Fermi level and become occupied as the concentration of Sn
increases. Thus the Mn spin moment increases practically linearly
from 2.6 to 3.2 $\mu_B$. The higher polarization of the Mn $d$
states polarizes also the Co bands since they form a common
majority band and the Co moment increases by 0.2 $\mu_B$.

\begin{figure}[t]
\includegraphics[scale=0.47]{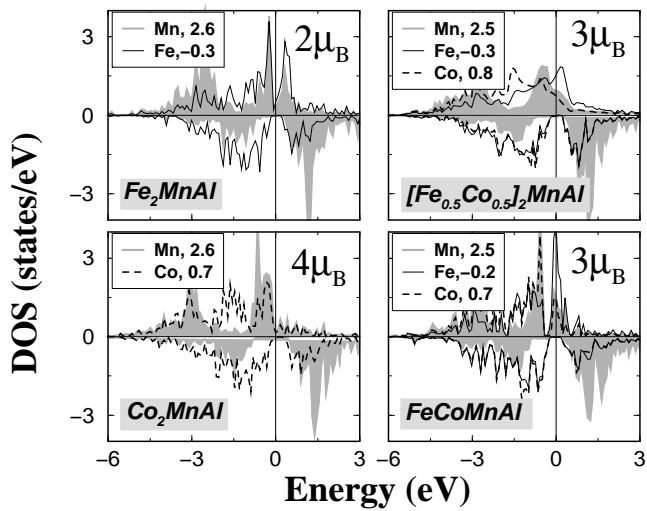}
 \caption{\label{fig3}Atom projected DOS in the case that
  both X sites are occupied by Co and Fe with
  50\% probability (system is denoted by
  [Fe$_{0.5}$Co$_{0.5}$]$_2$MnAl)
  and when one sublattice is occupied exclusively by Fe and the other by Co
 atoms (system is denoted by FeCoMnAl). All compounds are half-metals.
  The numbers in the legends are the
  atomic spin moments in $\mu_B$ and the larger ones are the total spin
  moments.}
\end{figure}

The last case I will discuss is when I mix the high-valent
transition metal atoms. The first family is the
[Fe$_{1-x}$Co$_x$]$_2$MnAl, where the total spin moment increases
from 2 to 4 $\mu_B$ with the concentration. The Al moment stays
small and is negligible while the Mn moment is 2.6 $\mu_B$ for all
concentrations. The Fe moment varies form $-$0.32 to $-$0.26
$\mu_B$ and the Co one from 0.63 to 0.75 $\mu_B$. Thus the atomic
spin moments change little and the increase in the total spin
moment arises exclusively from the substitution of Fe by Co. It is
also interesting to compare [Fe$_{0.5}$Co$_{0.5}$]$_2$MnAl with
$\mathrm{FeCoMnAl}$. In the latter compound one of the sublattices
is occupied exclusively by Fe atoms and the other by Co ones. Both
systems are HMs as can be seen in Fig. \ref{fig3} and the total
spin moment is 3 $\mu_B$. Both compounds show similar atomic spin
moments, thus the exact position of the Fe and Co atoms is not so
relevant for the magnetic properties, which are controlled mainly
by the concentration of the chemical elements. If I compare the
DOSs presented in Fig. \ref{fig3}, I see that the atom projected
ones have the same characteristics although the ones for FeCoMnAl
are more spiky. The main difference is a small majority gap  just
below the Fermi level in FeCoMnAl which is washed out for
[Fe$_{0.5}$Co$_{0.5}$]$_2$MnAl. This gap is the signature of the
order. \\ \noindent The second family of compounds is the
[Rh$_{1-x}$Co$_x$]$_2$MnAl. Rh and Co are isoelectronic elements
and the total spin moment is 4 $\mu_B$ for all concentrations.
Also RhCoMnAl is a HM with magnetic properties similar to
[Rh$_{0.5}$Co$_{0.5}$]$_2$MnAl as it was the case for the
compounds containing Fe. Since Rh moment is much smaller than the
Co one,  the Mn moment has to increase considerably, from 2.6 to
3.4 $\mu_B$, with the increase of Rh concentration to keep the
total spin moment constant,.

A special case uder study is the [Mn$_{1-x}$Co$_x$]$_2$VAl family.
For this compound the spin moment changes from $-$2 to 2 $\mu_B$
with the concentration $x$. As shown in Ref.
\onlinecite{Galanakis02b} the Mn and V atoms in Mn$_2$VAl  are
antiferromagnetically coupled and V carries a large spin moment,
while in Co$_2$VAl the Co and V atoms are ferromagnetically
coupled and the moment is mainly carried by the Co atoms. For
$x$=0.5 there is the case of a compound with zero total spin
moment but with magnetic constituents; $m^\mathrm{Mn}=-0.47\mu_B$,
$m^\mathrm{Co}=0.20\mu_B$ and $m^\mathrm{V}=0.23\mu_B$. The Mn
spin moment decreases from $-$1.45 $\mu_B$ to -0.35 with the
concentration $x$ while the Co spin moment increases from 0.2 to
0.95 $\mu_B$ and the V one decreases from 0.8 to 0.16 $\mu_B$.

In this contribution I expanded the study already performed for
the ordered Heusler alloys to cover also the quaternary Heusler
compounds like Co$_2$[Cr$_{1-x}$Mn$_x$]Al,
Co$_2$Mn[Al$_{1-x}$Sn$_x$] and
 [Fe$_{1-x}$Co$_x$]$_2$MnAl. Using the
Korringa-Kohn-Rostoker method in the coherent potential
approximation I have shown that independently from which site is
disordered, all compounds are half-metals and the total spin
moment $M_t$ scales linearly with the total number of valence
electrons $Z_t$ following the rule $M_t=Z_t-24$, thus $M_t$ scales
also linearly with the concentration $x$. Depending of the family
under study the atomic spin moments change in a way that the above
rule, known also as the Slater-Pauling behavior, is obeyed.
Finally an interesting case is [Mn$_{1-x}$Co$_x$]$_2$VAl where the
moment scales linearly from $-2$ $\mu_B$ to 2 $\mu_B$ and for
$x=0.5$ the total spin moment is zero but the constituents are
magnetic.

\begin{acknowledgments}
 Financial support from the RT Network of Computational Magnetoelectronics
 (Contract RTN1-1999-00145) of the European Commission is greatfully acknowledged.
 The author would like to thank Professor P. H. Dederichs for stimulating discussions
  and Dr. K. Sato for his help during the calculations.
\end{acknowledgments}

\end{document}